\documentclass[aps,reprint,showpacs,floatfix,citeautoscript,longbibliography,superscriptaddress]{revtex4-2}

\usepackage{graphicx}
\usepackage{dcolumn}
\usepackage{bm}
\usepackage[colorlinks=true, linkcolor=blue, citecolor=blue, urlcolor=blue, linktoc=page, bookmarks=false]{hyperref} 

\usepackage{graphicx}
\usepackage{dcolumn}
\usepackage{bm}

\usepackage[utf8]{inputenc}
\usepackage[T1]{fontenc}
\usepackage{mathptmx}
\usepackage{etoolbox}

\makeatletter
\def\@email#1#2{%
 \endgroup
 \patchcmd{\titleblock@produce}
  {\frontmatter@RRAPformat}
  {\frontmatter@RRAPformat{\produce@RRAP{*#1\href{mailto:#2}{#2}}}\frontmatter@RRAPformat}
  {}{}
}%
\makeatother
\begin{document}

\preprint{APS/123-QED}

\title{Bayesian Reasoning Enabled by Spin-Orbit Torque Magnetic Tunnel Junctions}
\author{Yingqian Xu}
 \affiliation{Beijing National Laboratory for Condensed Matter Physics, Institute of Physics, University of Chinese Academy of Sciences, Chinese Academy of Sciences, Beijing 100190, China}
 \affiliation{Center of Materials Science and Optoelectronics Engineering, University of Chinese Academy of Sciences, Beijing 100049, China}
\author{Xiaohan Li}
\affiliation{Beijing National Laboratory for Condensed Matter Physics, Institute of Physics, University of Chinese Academy of Sciences, Chinese Academy of Sciences, Beijing 100190, China}

\author{Caihua Wan}
\affiliation{Beijing National Laboratory for Condensed Matter Physics, Institute of Physics, University of Chinese Academy of Sciences, Chinese Academy of Sciences, Beijing 100190, China}
\affiliation{Songshan Lake Materials Laboratory, Dongguan, Guangdong 523808, China}

\author{Ran Zhang}
\affiliation{Beijing National Laboratory for Condensed Matter Physics, Institute of Physics, University of Chinese Academy of Sciences, Chinese Academy of Sciences, Beijing 100190, China}
\author{Bin He}
\affiliation{Physical Science and Engineering Division, King Abdullah University of Science and Technology (KAUST), Thuwal, 23955–6900, Saudi Arabia}
\author{Shiqiang Liu}
\affiliation{Beijing National Laboratory for Condensed Matter Physics, Institute of Physics, University of Chinese Academy of Sciences, Chinese Academy of Sciences, Beijing 100190, China}
\author{Jihao Xia}
\affiliation{Beijing National Laboratory for Condensed Matter Physics, Institute of Physics, University of Chinese Academy of Sciences, Chinese Academy of Sciences, Beijing 100190, China}
\author{Dehao Kong}
\affiliation{Beijing National Laboratory for Condensed Matter Physics, Institute of Physics, University of Chinese Academy of Sciences, Chinese Academy of Sciences, Beijing 100190, China}
\author{Shilong Xiong}
\affiliation{Beijing National Laboratory for Condensed Matter Physics, Institute of Physics, University of Chinese Academy of Sciences, Chinese Academy of Sciences, Beijing 100190, China}
\author{Guoqiang Yu}
\affiliation{Beijing National Laboratory for Condensed Matter Physics, Institute of Physics, University of Chinese Academy of Sciences, Chinese Academy of Sciences, Beijing 100190, China}
\affiliation{Songshan Lake Materials Laboratory, Dongguan, Guangdong 523808, China}
\author{Xiufeng Han}
\email[Authors to whom correspondence should be addressed: ]{[Xiufeng Han, xfhan@iphy.ac.cn; Caihua Wan, wancaihua@iphy.ac.cn]}
\affiliation{Beijing National Laboratory for Condensed Matter Physics, Institute of Physics, University of Chinese Academy of Sciences, Chinese Academy of Sciences, Beijing 100190, China}
\affiliation{Center of Materials Science and Optoelectronics Engineering, University of Chinese Academy of Sciences, Beijing 100049, China}
\affiliation{Songshan Lake Materials Laboratory, Dongguan, Guangdong 523808, China}

\date{\today}

\begin{abstract}
Bayesian networks play an increasingly important role in data mining, inference, and reasoning with the rapid development of artificial intelligence. In this paper, we present proof-of-concept experiments demonstrating the use of spin-orbit torque magnetic tunnel junctions (SOT-MTJs) in Bayesian network reasoning. Not only can the target probability distribution function (PDF) of a Bayesian network be precisely formulated by a conditional probability table as usual but also quantitatively parameterized by a probabilistic forward propagating neuron network. Moreover, the parameters of the network can also approach the optimum through a simple point-by-point training algorithm, by leveraging which we do not need to memorize all historical data nor statistically summarize conditional probabilities behind them, significantly improving storage efficiency and economizing data pretreatment. Furthermore, we developed a simple medical diagnostic system using the SOT-MTJ as a random number generator and sampler, showcasing the application of SOT-MTJ-based Bayesian reasoning. This SOT-MTJ-based Bayesian reasoning shows great promise in the field of artificial probabilistic neural network, broadening the scope of spintronic device applications and providing an efficient and low-storage solution for complex reasoning tasks.
\end{abstract}

\maketitle

%

\section{Introduction}
The rapid development of artificial intelligence (AI) over the past few decades has been nourished by advancements in machine learning algorithms, increased computational power, and availability of vast amounts of data\cite{ref1}, which has in turn revolutionized numerous fields including but not limited to medical science and healthcare, information technologies, finance, transportation, and more. This regenerative feedback between AI and its applications leads to a further explosive growth of data and expansion of model scales, which calls for a paradigm shift toward efficient and speedy computing and memory technologies, especially, advanced algorithms and emerging AI hardware enabled by nonvolatile memories\cite{ref2}.

In this aspect, the emerging memory technologies, such as magnetic random-access memories\cite{ref3}, ferroelectric random-access memories\cite{ref4}, resistive random-access memories\cite{ref5,ref6} and phase-change random-access memories\cite{ref7}, have been implemented to accelerate AI computing, for instance, the matrix multiplication\cite{ref8}. Thanks to their high energy-efficiency, fast speed, long endurance, and versatile functionalities, spintronic devices based on spin-orbit torques as one prominent example among emerging memories, have shown great potential in the aspect of hardware-accelerated true random number generation (TRNG)\cite{ref9,ref10,ref11,ref12,ref13,ref14,ref15,ref16,ref17,ref18} besides of the matrix multiplication. For instance, the high quality true random number generators with stable and reconfigurable probability-tunability have been demonstrated using SOT-MTJs \cite{ref19,ref20,ref21}. Worth noting, the TRNG task is especially impactful for probabilistic neuron networks aiming at optimization, learning, generation, reasoning and inference\cite{ref22}. The optimization task of an MTJ-based neuron network has been experimentally demonstrated for the integer factorization\cite{ref23,ref24} or for the traveling salesman problem with the non-deterministic polynomial hardness\cite{ref25,ref26}. The cross-modal learning and generation has also been realized in a SOT-MTJ-based restricted Boltzmann machine\cite{ref23,ref27}. However, the reasoning and inference task of probabilistic neuron networks accelerated by spintronic devices is still rare\cite{ref22} and to be actualized and enriched.

Bayesian networks, a category of directed graph models, excel in expressing probabilistic causal relationships among a set of random variables. Their ability to incorporate prior knowledge and update beliefs with new evidence makes Bayesian networks particularly powerful frameworks for reasoning and inference\cite{ref28,ref29}. Any directed edge in such a Bayesian network denotes a causal relationship from a parent node (one cause of an event or the start of the edge) to a child node (one outcome of the event or the terminal of the edge). Owing to their excellence in encoding causal relations, these Bayesian networks have been widely used in prediction, anomaly detection, diagnostics, and decision-making under uncertainty\cite{ref30,ref31,ref32,ref33}.

However, due to complexity of the real world, one outcome (reason) can result from (in) different reasons (outcomes) and moreover one outcome of a previous event can even cascade a subsequent event as its starting reason. Thus, Bayesian networks can be deeply multi-leveled and contain a large number of nodes in practice. It is thus not surprising that building of a Bayesian network is burdensome. Moreover, a Bayesian network only offers us a logic framework in concept; to implement it in encoding practical causal relations, we need to organize it in the form of a conditional probability table (CPT) as elaborated in Ref\cite{ref19} in which massive historical data should be stored and then statistically counted into many conditional probabilities. Or the joint probability-distribution-function (PDF) of the whole system (considering all random variables) should be stable and already known. The translation between the stable PDF and CPT will be introduced in details below. Nevertheless, both methods are memory-intensive and historical data should be properly structured in the format of a CPT or PDF. Here we develop a simple point-by-point training algorithm that do not rely on any structured nor historical data but only on the ‘present’ observation point to effectively parameterize and train a Bayesian network. The automatically trained network, though point-by-point, can still quantitatively reproduce the overall PDF of all the historical data and accurately describe the causal relationship between parent-child pairs in the Bayesian network. This algorithm also enables dynamically fine-tuning network parameters according to new coming data, if given, to keep real-time correctness of a model. Furthermore, we have also shown spin-orbit torque magnetic tunnel junctions (SOT-MTJs) were qualified competent to act as probabilistic samplers, which paves a feasible avenue for hardware trained Bayesian networks.

\section{Experiments}
\begin{figure*}[htbp]
    \includegraphics[width=16cm]{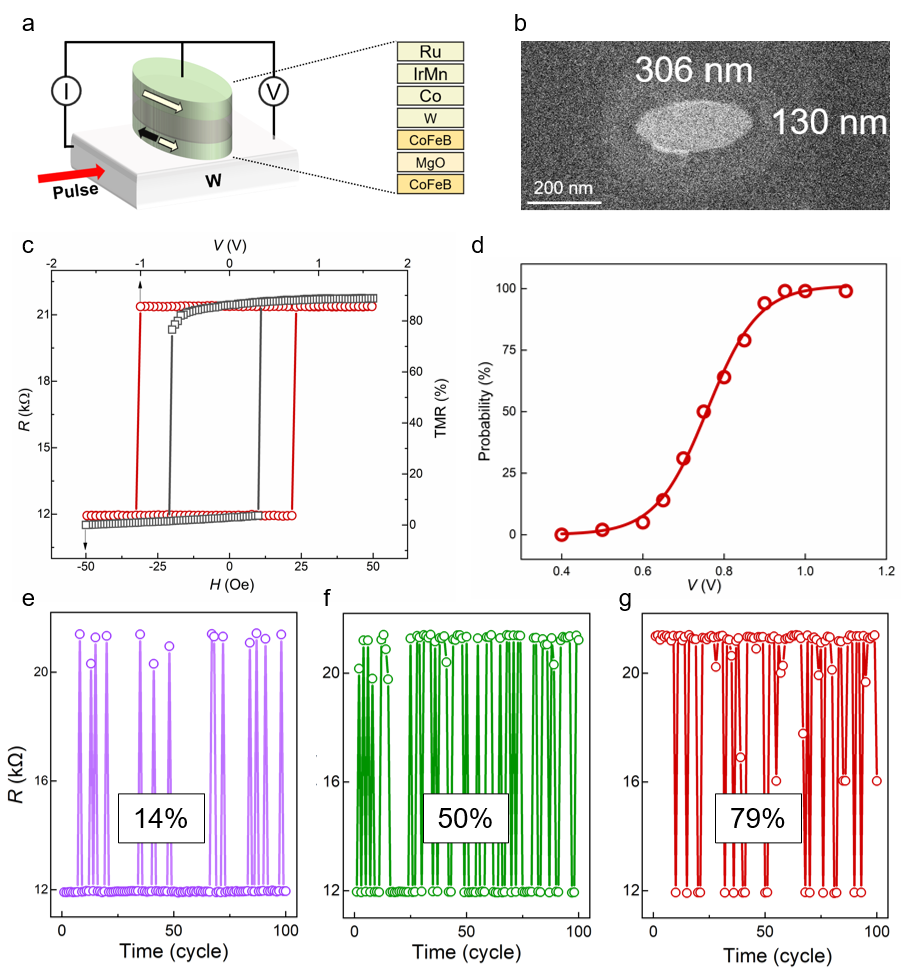}
    \caption{Characterization of Y-Type SOT-MTJs. (a) Structure of Y-Type SOT-MTJ. (b) SEM image of an MTJ. (c) \emph{R-H} loop of the Y-type SOT-MTJ obtained with an in-plane field along the easy axis. Field-free switching of the free layer induced by a 50 ns voltage pulse. (d) Relationship between probability and driven voltage. (e - h) Results obtained from continuous measurement under driven voltages of 0.9 V (e), 1.05 V (f), 1.2 V (g).}
    \label{fig:1}
\end{figure*}
As shown in Fig.\ref{fig:1}a, the MTJ stack consists of W(3)/CoFeB(1.4)/MgO(1.5)/CoFeB(3) /W(0.4)/Co(2.7)/IrMn(10)/Ru(4 nm). The numbers in parentheses indicate nominal thicknesses in nanometers. The stack was deposited in a magnetron sputtering system (ULVAC) at room temperature and then annealed in vacuum at 380 ℃ to obtain in-plane uniaxial magnetic anisotropy. After annealing, it was patterned into an ellipse using electron-beam lithography (EBL), reactive ion etching (RIE) and ion beam etching (IBE) as described in Ref\cite{ref34}. The resistance of SOT-MTJs was measured using the four-probe method with a Keithley 2400 source meter and a Keithley 2182 nanovoltmeter, and current pulses were applied to the write line by an Agilent 81104A pulse generator. Figure\ref{fig:1}b depicts a typical scanning electron microscope (SEM) image of an SOT-MTJ device (the top view). The device is calibrated as an ellipse of 130 nm×306 nm, exhibiting in-plane uniaxial magnetic anisotropy along its long axis. The resistance of SOT-MTJs depends on the relative magnetization orientation of the free layer with respect to the reference layer. The parallel (antiparallel) configuration corresponds to the low (high) resistance state in our case. Switching between the two states is achievable by a magnetic field H or simply a current/voltage (\emph{V}) pulse in the field-free condition (Fig.\ref{fig:1}c). Figure \ref{fig:1}c shows the dependence of MTJ resistance on \emph{H} along the easy axis,where the magnetic field is generated by a Helmholtz coil (3D magnetic field probe station, East Changing Technologies, China). Its TMR ratio is \textasciitilde 100$\%$, indicating high quality of the MTJ stack. The magnetization of the free layer can be switched by a 50 ns current pulse flowing in the write line with $H = 0$ as well (Fig.\ref{fig:1}c).

To obtain the switching probability, a 50 ns pulse voltage of -1.1 V is initially applied to reset the MTJ to its low resistance state. Subsequently, a pulse voltage with a specific amplitude \emph{V} is applied to attempt SOT-MTJ switching. The resistance is then monitored after switching. This procedure is referred as a reset-sampling operation circle. Experiencing a reset-sampling circle, a MTJ can transit to the high resistance state (random number = 1) or remain in the low resistance state (random number = 0). Figure \ref{fig:1}d depicts the dependence of the switching probability \emph{P} on the write voltage. Here each point was statistically calculated from 100 independent reset-sampling cycles. The data well fit a sigmoid function as marked by the red curve. As a result, \emph{P} can be continuously and precisely tuned by \emph{V}. Figures.\ref{fig:1}e-g demonstrate the resistance state of a SOT-MTJ device at the voltages of 0.65 V, 0.75 V and 0.85V, corresponding to \emph{P} of 14\%, 50\% and 79\%, respectively.

Till now, we have demonstrated that SOT-MTJs can function ideally as a \emph{P}-tunable TRNG. Hereafter, we employ SOT-MTJs as a decision maker/generator in Bayesian networks.

\section{Results and Discussions}
\begin{figure*}[htbp]
    \includegraphics[width=16cm]{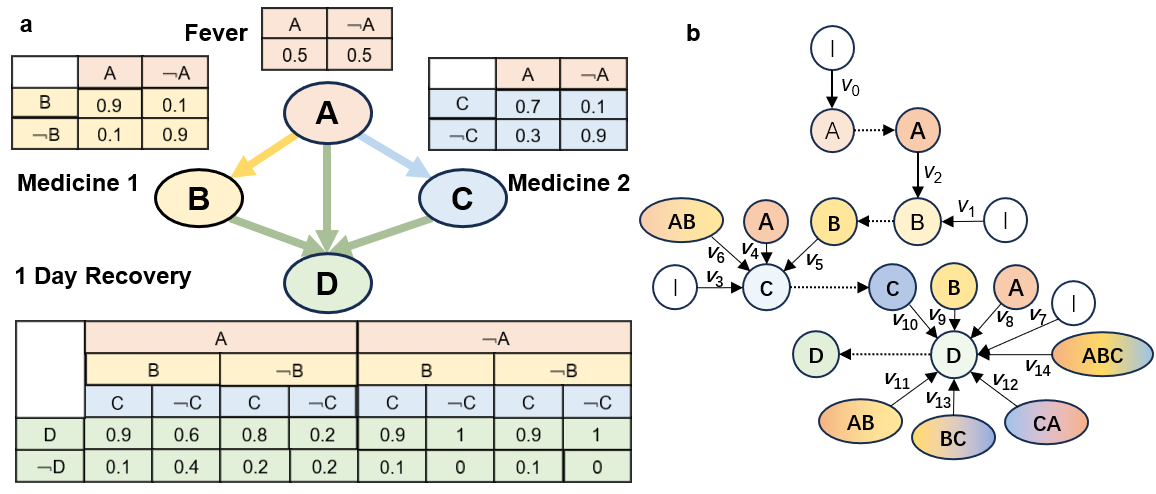}
    \caption{The network structure for Bayesian network reasoning. (a) The Bayesian network and conditional probability table (CPT) for generating samples or medical cases. (b) The network used to calculate the probability density function (PDF) corresponding to $v_{i}$. }
    \label{fig:2}
\end{figure*}
In the following, we first demonstrate that a SOT-MTJ-based Bayesian network can generate random numbers accordingly to any desired probability-distribution-functions (PDF) once edges of the network is properly weighted. For instance, we built a 4-node Bayesian network to demonstrate the PDF-configurable TRNG. Each node ($A$, $B$, $C$ and $D$) represents a bit of a four-digital binary number \emph{N} as shown in Eq.(\ref{eq:1}). The task of encoding a desired PDF \emph{P(N)} is then reduced into another one, finding a suitable causal relationship among binary random variables \emph{A-D} which corresponds to the targeted \emph{P(N)}. Fortunately, the above two tasks are proven mathematically equivalent and the network parameters can be straightforwardly derived from the desired PDF. Here, we detail the transformation procedure from the desired PDF into the network weights or vice versa.
\begin{equation}
  N = {2}^{3}A + {2}^{2}B + {2}^{1}C + {2}^{0}D 
\label{eq:1}
\end{equation}

Figure \ref{fig:2}a shows the Bayesian network to encode the \emph{P(N)}. As mentioned above, it contains 4 nodes corresponding to the 4-digits of \emph{N}. Due to the decreasing pre-factors of \emph{A-D} in Eq.(\ref{eq:1}), their influences to \emph{N} also attenuate. Therefore, \emph{A (B)} acts as the parent node of \emph{B-D (C-D)}. So does \emph{C} to \emph{D}. It means the probability of \emph{B} = 1 is determined by the value of \emph{A} (after probabilistically sampling \emph{A}), the probability of \emph{C} = 1 is further decided by both values of \emph{A} and \emph{B} (after sampling \emph{B}) and so on. This scenario can be further conveniently encoded by a forwardly propagating neuron network with the binary random variables \emph{A-D} as well as their probabilistic sampling operations, which represents one invention of this work. As shown below, this forward neuron network with random variables offers another parameterization method for the Bayesian network, from which the ideal PDF of the network can be directly formulated from its network parameters. By comparing this ideal one with the experimental one (the stable PDF in experiment or even every single sampling point), one can train the forward neuron network and finally allow it to output samples adhering to the experimental PDF as desired.

As illustrated in Fig.\ref{fig:2}b, the network consists of 4 layers and each layer has to probabilistically sample 1 node. Here the collapse from a random number to its sampling result is denoted by a dashed line. For Node $A$, the switching probability $p_A$ is given by $p_A=f_A (v_0)$, where $v_0$ corresponds to the weight of the edge connected to Node $A$ in the first layer and $I = 1$ is a constant node. The function $f_i (\cdot)$ represents the $V$-dependence of $P$ of the $i^{th}$ node, which is analog to the sigmoid function in Fig.\ref{fig:1}e. Node $B$ is the child node of Node $A$, so $p_B$ is influenced by the state of $A$: $p_B=p_{B|A}=f_B (v_1+Av_2)$. If $A = 0$, then $p_B=f_B (v_1)$; if $A = 1$, then $p_B=f_B (v_1+v_2)$. In this case, there are two edges (or two independent weights $v_1$ and $v_2$) connecting to Node $B$ in the $2^{nd}$ layer. Similarly, Node $C$ is the child node of Parents $A$ and $B$. $p_c$ is determined by the states of both $A$ and $B$ after sampling or $p_C=p_{C|AB}=f_C (v_3+Av_4+Bv_5+ABv_6)$. Here 4 weights are necessary. Especially, the term $v_6$ characterizes the joint act of $A$ and $B$ on $C$. Likewise, Node $D$ is a child of Nodes $A$-$C$, and its probability $p_D$ is given by $p_D=p_{D|ABC}=f_D (v_7+Av_8+Bv_9+Cv_{10}+ABv_{11}+ACv_{12}+BCv_{13}+ABCv_{14}$, 8 parameters being indispensable to represent their independent or combined effects on $D$.
\begin{equation}
  p(\bm{x}) = \prod_{k=1}^{K} p(x_k|\bm{x}_{{\pi}_{k}} ) 
\label{eq:2}
\end{equation}

As shown above, the final probability of each node, 0 or 1, is determined by the sampled state of its parent nodes. According to the Bayesian theory, the joint probability distribution of all variables can be expressed as the product of the conditional probabilities of each random variable in Eq.(\ref{eq:2}). Therefore, the joint PDF of Nodes $A$-$D$ can be expressed as $p_{ABCD}=p_{A} \times p_{B|A} \times p_{C|AB} \times p_{D|ABC}=f_{A} \times f_{B} \times f_{C} \times f_{D}$. Actually, $p_{ABCD}=P(N)$ with $N = 0, 1, 2, …, 15$ if one recalls Eq.(\ref{eq:1}). Then we have $2^{4}=16$ equations in total. Nevertheless, $\sum_0^{15}P(N)=1$, we have only 15 independent equations. Using these independent equations, we can straightforwardly calculate the value of $v_{l} (l = 0, 1, …, 14)$. The relation between $v_{l} (l = 0, 1, …, 14)$ and $P(N)$ can thus be described by a $15 \times 15$ matrix.

\begin{figure*}
    \includegraphics[width=16cm]{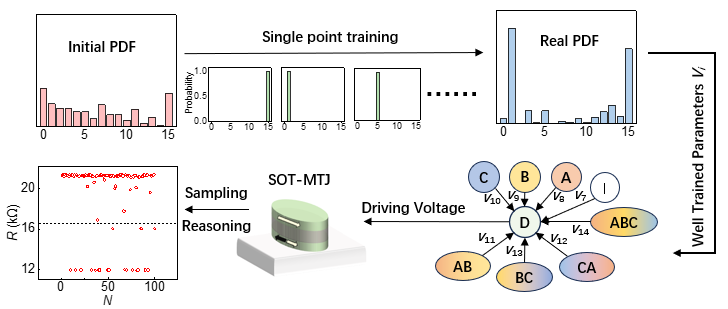}
    \caption{The reasoning process of the Bayesian network utilizing the SOT-MTJ. }
    \label{fig:3}
\end{figure*}

\begin{figure*}
    \includegraphics[width=16cm]{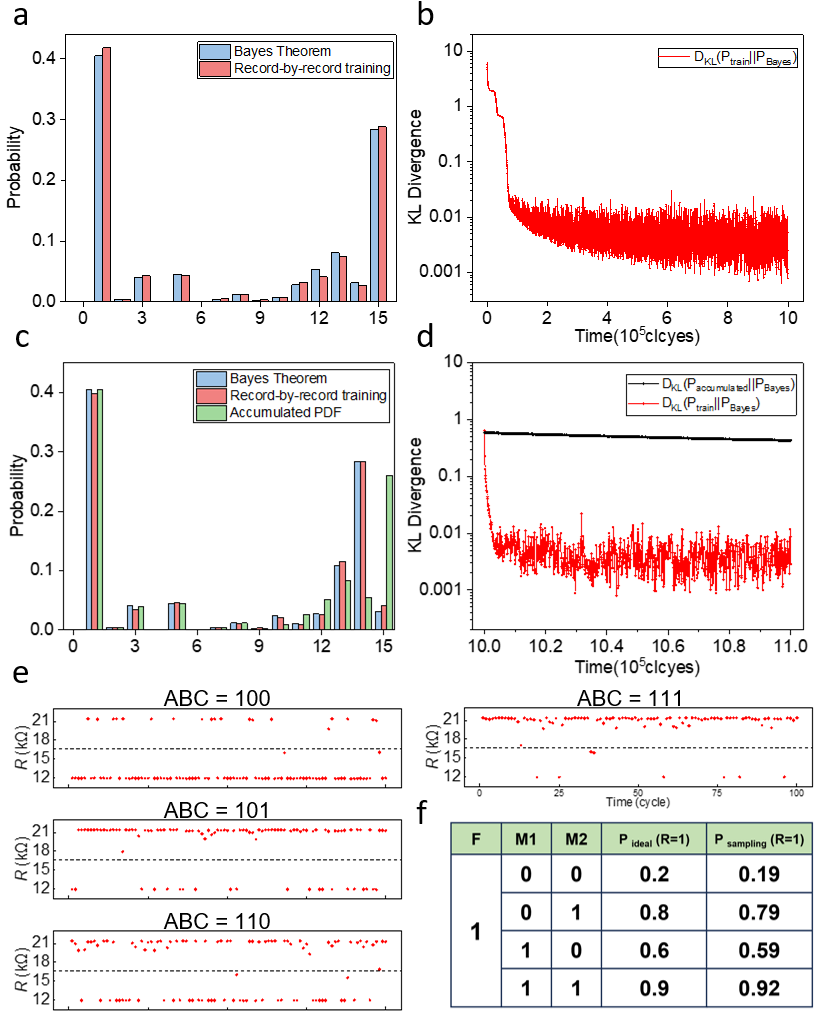}
    \caption{(a) Histogram of the state probability of our single-point training algorithm compared to the Bayesian theory. (b) The K-L divergence between our single-point training algorithm and the Bayesian theory decreases with the increase in training cycles. (c) Histogram of the state probability of our single-point training algorithm compared to the Bayesian theory and the accumulated PDF after $10^{5}$ new records. (d) The K-L divergence of our single-point training algorithm (red) or the accumulated PDF (black) with the Bayesian theory from the $10^{5}$ new records. The experimental sampling results (e) and statistical results of generating 1 (f) in the cases of $ABC$ = 100, 101, 110, and 111. }
    \label{fig:4}
\end{figure*}

Up to now, we have derived that the PDF $P(N)$ and the network parameters $v_l$ can be mutually transformed. If $P(N)$ is already known, we can directly obtain $v_l$ accordingly. Reversely, if the network parameters $v_l$ are given, we can then obtain the corresponding ideal $P_{cal}(N)$ of the network. Interestingly, if we compare $P_{cal}(N)$ calculated from $v_{l}$ with the experimental $P_{exp}(N)$, we can be informed how to further adjust $v_{l}$ to minimize the K-L distance between $P_{cal}$ and $P_{exp}$. Following this idea, the network can be trained to learn $P_{exp}(N)$. More crucially, we do not have to statistically count a complete $P_{exp}(N)$ in practice. Instead, we just used every single point $N$ iteratively to train the network well. In this case, we actually let $P_{exp}(X=N) = \delta(N)$ with the probability of $X = N (X \neq N)$ being 100\% (0).

Next, we integrate this idea with an algorithm to design an automatic medical diagnostic system. As shown in Fig.\ref{fig:2}a, we map the 4 nodes to 4 events. Nodes $A$-$D$ correspond to Fever, Medicine 1, Medicine 2 and Recovery within one day, respectively. Obviously, Fever is the original reason for the other three events, Medicine 1 and 2 are the treatments to the Fever and Recovery (or not) is the final outcome of the Fever and the treatments. Using the CPT in Fig.\ref{fig:2}a, we randomly generate a dataset of $10^{6}$ samples. Each sample is encoded by a number $N$ between 0 and 15, corresponding to a medical record. For example, Sample ‘15’ in the decimal system or ‘1111’ in the binary digitals corresponds to ‘$A = 1, B = 1, C = 1, D = 1$’, implying that a Fever patient after taking Medicine 1 and Medicine 2 together Recovered within one day.
\begin{equation}
D_{KL}(P||P^{cal})=\sum_{ijmn}P_{ijmn}log\frac{P_{ijmn}}{P_{ijmn}^{cal}}
\label{eq:3}
\end{equation}
\begin{equation}
\frac{\partial D_{KL}(P||P^{cal})}{\partial v_{l}}=\sum_{ijmn}(-\frac{P_{ijmn}}{P_{ijmn}^{cal}}\frac{\partial P_{ijmn}^{cal}}{\partial v_{l}})
\label{eq:4}
\end{equation}

Hereafter, we detail the reasoning process of the medical diagnostic system based on the Bayesian network, as shown in Fig.\ref{fig:3}. We first initialize $v_{l}$ $(l= 0 \sim 14)$ to 0.5 V and calculate the corresponding PDF. Here the network weights are encoded by the write voltage $V$ of SOT-MTJ devices because this parameter directly decides their switching probability and is also continuously controllable. Then we randomly select a medical record from the dataset mentioned above one-by-one and treat the record as a delta PDF $P=\delta(N)$ for this point. And then we attempt to adjust $v_{l}$ as well as the corresponding computed PDF $P^{cal}$ to approach every delta PDF defined by each record and minimize their K-L distance as defined in Eq.(\ref{eq:3}). The training method is explicitly described below: Using Eq.(\ref{eq:4}), we calculate the partial derivative of the K-L Divergence with respect to $v_{l}$ $(l=0 \sim 14)$. Inspired with the gradient descent algorithm, $v_{l}$ is updated as following $v_{l}=v_{l}- \alpha \partial D_{KL}(P||P^{cal})/\partial v_{l}$. Here, $\alpha =5 \times 10^{-5}$ is the learning rate. These training steps are iteratively repeated, once for one medical record. As shown in Fig.\ref{fig:4}a-b, after training with $10^{6}$ medical records, the obtained PDF aligns well with the result from the Bayesian theory. The K-L divergence is used to describe the difference between the two distributions. As the update cycles increase, the K-L divergence between the PDF trained by our single-point training method and the PDF from the Bayesian theory decreases gradually, implying the convergence and the effectiveness of the training method.

Unlike statistically averaged approaches, our training process is point-by-point. It means the network is automatically trained every time a record emerges, and the training weights are stored and refreshed in the network parameters $v_{l}$ $(l=0\sim14)$ in real time, neither need to save massive historical data nor statistical count of CPT from them. What we need are just one-by-one raw data without pretreatment and without necessity of memorizing them after use. Apparently, the training process of the Bayesian network based on our algorithm can save storage space and statistic cost. Moreover, this algorithm permits the network dynamically tuning its parameters to accommodate sudden changes of the experimental PDF in a real-time fashion. For example, we still imagine the above ‘virtual’ disease that causes fever. If a gene mutation occurs to the causative virus of the disease, the effectiveness of Medicine 1 is sharply reduced from $P_{ideal}(1)$ = 0.8 to 0.3 while that of Medicine 2 is mildly increased from $P_{ideal}(2)$ = 0.6 to 0.8. Even worse, the joint act of Medicine 1 and 2 leads to serious counter reaction and the resultant recovery rate is thus reduced from 0.9 to 0.1. $10^{5}$ new records are stochastically generated from this suddenly changed PDF to mimic the influence of the gene mutation. In this situation, if we still use the PDF accumulated from the whole historical data, $10^{6}$ old records + $10^{5}$ new records, the network with its parameters computed directly from the PDF apparently predicts a diverse distribution from the suddenly changed one (Fig.\ref{fig:4}c-d). However, by persisting in updating the parameters by the record-by-record training algorithm, the network can soon feel the sudden change in PDF and quickly reproduce it correctly as shown in Fig.\ref{fig:4}c-d, manifesting adaptability and powerfulness of this point-by-point training protocol.

According to the stable PDF obtained after training, we calculate the corresponding network parameters $v_{l}$ $(l=0\sim14)$. Using these parameters, we implement a simple but automatic medical diagnostic system with the PDF before the gene mutation. As shown in Fig.\ref{fig:4}e, the amplitude of the write voltage applied to Node $D$ is determined by the state of Nodes $A$-$C$. When the states of Nodes $A$-$C$ are fixed, we apply the corresponding writing voltage to the SOT-MTJ. After 100 cycles of reset-sampling operations, we statistically count the probability of $D=1$, which corresponds to the probability that patients recover within one day after a certain treatment. We study the cases of $A$ = 1, $BC$ = 00, 01, 10 and 11, and compare their statistical sampling results in Fig.\ref{fig:4}f. This experiment reflects the probability that a patient with a fever will recover within one day after taking different recipes. The results indicate that Medicine 1 is more effective than Medicine 2, and the concomitant use of Medicines 1 and 2 produces an even higher recovery rate. By comparing the recovery probabilities of various recipes, the system can automatically recommend the best one, thus implementing reasoning and decision-making tasks.
\section{conclusion}
In conclusion, we have conducted proof-of-concept experiments demonstrating the Bayesian network reasoning utilizing the SOT-MTJ. By integrating the $P-V$ sigmoid function of the SOT-MTJ into a 4-node Bayesian network, we accurately calculated the PDF using the network parameters $v_{l}$   $(l=0\sim14)$, corresponding to the writing voltage applied to the SOT-MTJ. We then developed a point-by-point training algorithm to stabilize the parameters $v_{l}$ as well as the Bayesian network dynamically. Compared to the statistical method, this algorithm does not require storing all historical data, significantly reducing the needed storage space and also increasing adaptability. After training the network, we compared the statistical results of sampling under different node states, demonstrating the SOT-MTJ functioned properly as a reasoning maker in a simple automatic medical diagnostic system. This SOT-MTJ-based Bayesian network for reasoning has great potential in the field of artificial neural networks, significantly expanding the application range of spintronic and SOT-MTJ devices.

\begin{acknowledgments}
This work was supported by the National Key Research and Development Program of China (MOST) (Grant No. 2022YFA1402800), the National Natural Science Foundation of China (NSFC) (Grant No. 12134017 and 12374131), the Strategic Priority Research Program (B) of Chinese Academy of Sciences (CAS) (Grant Nos. XDB33000000). C. H. Wan appreciates financial support from the Youth Innovation Promotion Association, CAS (Grant No. 2020008).
\end{acknowledgments}

\section*{Data Availability Statement}
The data that support the findings of this study are available from the corresponding authors upon reasonable request.

\section{References}
\nocite{*}
\bibliography{reasoning}

\end{document}